\documentstyle[preprint,aps]{revtex}
\draft

\begin{document}


\preprint{YITP-98-30, gr-qc/9805039}

\title{Comments on entanglement entropy}
\author{Shinji Mukohyama}
\address{
Yukawa Institute for Theoretical Physics, 
Kyoto University \\
Kyoto 606-8502, Japan
}
\date{\today}

\maketitle


\begin{abstract} 

A new interpretation of entanglement entropy is proposed:
entanglement entropy of a pure state with respect to a division of a 
Hilbert space into two subspaces $1$ and $2$ is an amount of
information, which can be transmitted through $1$ and $2$ from a
system interacting with $1$ to another system interacting with
$2$. The transmission medium is quantum entanglement between $1$
and $2$. In order to support the interpretation, suggestive arguments
are given: variational principles in entanglement thermodynamics and
quantum teleportation. It is shown that a quantum state having 
maximal entanglement entropy plays an important role in quantum
teleportation. Hence, the entanglement entropy is, in some sense, an
index of efficiency of quantum teleportation. 
Finally, implications for the information loss problem and Hawking
radiation are discussed. 

\end{abstract}


\newpage


\section{Introduction}
	\label{sec:intro}

It is believed that the well-known analogy between the theory of black 
holes and thermodynamics, which is called black hole thermodynamics, is 
universal and that there should be a deep physics in its origin. 
In particular, since its first introduction by Bekenstein~\cite{Bekenstein} 
as a quantity proportional to horizon area, black hole entropy 
has been one of the hottest topics in black hole physics. 
The black hole entropy is called the Bekenstein-Hawking entropy because 
the proportionality coefficient was determined to be $1/4$ by Hawking's
discovery of thermal radiation from a black hole~\cite{Hawking1975}. 
It is well known that the black hole entropy has 
various properties similar to those of thermodynamical entropy. For 
example, the sum of the black hole entropy and the entropy of matter
outside the black hole does not decrease. This fact is called the
generalized second law and was proved for a quasi-stationary black 
hole~\cite{Frolov&Page,Mukohyama-GSL}. Moreover, the black hole entropy 
can be used to judge whether a black hole solution is stable or 
unstable~\cite{Catastrophy}. In fact, it can be used as a potential 
function in catastrophe theory and results of the catastrophe theory
coincide with those from a linear perturbation analysis. Hence, there 
are so many similarities
between black hole entropy and thermodynamical entropy that we 
expect that the former has a statistical origin, as the latter has.

Recently a microscopic derivation of the black hole entropy was given in 
superstring theory~\cite{GSW} by using the so-called D-brane 
technology~\cite{Polchinski}. In this approach the black hole 
entropy is identified with the logarithm of the number of states of 
massless strings attached to D-branes, with D-brane configuration 
and total momentum of the strings along a compactified direction fixed 
to be consistent with the corresponding black hole~\cite{D-brane}. The 
analysis along this line was extended to the so-called 
M-theory~\cite{Schwarz}. In particular, by using a conjectured 
correspondence (the Matrix theory) between the M-theory in the 
infinite momentum frame and a $10$-dimensional 
$U(N)$ supersymmetric Yang-Mills theory dimensionally reduced to 
$(0+1)$-dimension with $N\to\infty$~\cite{BFSS}, the black hole entropy 
was calculated by means of the Yang-Mills theory. The result gives the
correct Bekenstein-Hawking entropy for BPS black holes and their low
lying excitations~\cite{M-BPS}. Moreover, in Ref.~\cite{M-theory} the
black hole entropy of a Schwarzschild black hole was derived in the
Matrix theory up to a constant of order $1$. 
On the other hand, in loop quantum gravity~\cite{Rovelli}, black hole 
entropy was identified with the logarithm of the number of different 
spin-network states with the sum of eigenvalues of the area operator 
fixed~\cite{Loop}. The result coincides with the Bekenstein-Hawking entropy 
up to a constant of order $1$. 
It is evident that these derivations based on the candidate theories 
of quantum gravity depend strongly on details of the theories. In this 
sense, the success of the derivations can be considered as non-trivial 
consistency checks of the theories. However, it is believed 
that proportionality of the black hole entropy to horizon area 
is more universal and does not depend on details of the theory. 
Hence, one should be able to give a statistical or thermodynamical 
derivation of the black hole entropy, which does not depend on details 
of theory, while we are proceeding with theory-dependent derivations of 
it by using the well-established candidate theories of quantum gravity.

There were many attempts to explain the origin of the black hole entropy
besides the above theory-dependent approaches. 
For example, 
in Euclidean gravity the black hole entropy is associated with the
topology of an instanton which corresponds to a black 
hole~\cite{Topology}~\footnote{
The $1$-loop correction to the black hole entropy was also calculated
and compared with the brick wall model and the conical singularity 
method~\cite{1loop}.
}; 
Wald~\cite{Noether} defined the black hole entropy as a Noether charge 
associated with a bifurcating Killing horizon~\footnote{
Relations to the approach by Euclidean gravity was investigated in 
Ref.~\cite{Noe-Top}. 
};
'tHooft~\cite{Brick-Wall} identified the black hole entropy with the 
statistical entropy of a thermal gas of quantum particles with a 
mirror-like boundary just outside the horizon (the brick wall model);
Pretorius et al.~\cite{Shell} identified the black hole entropy with 
the thermodynamical entropy of a shell in thermal equilibrium with 
acceleration radiation due to the shell's gravity in the limit that
the shell forms a black hole. 
There remains another strong candidate for the statistical origin of 
the black hole entropy, called entanglement 
entropy~\cite{BKLS,Sred,Sent}. 
It is a statistical entropy measuring the information loss due to a 
spatial division of a system~\cite{BKLS}. 
The entanglement entropy is based only on the spatial division, and can 
be defined independently of the theory, although explicit calculations 
in the literature are dependent on the model employed. Moreover, as 
explained in the following argument, it is expected independently of the 
details of the theory that the entanglement entropy is proportional to 
the area of the boundary of the spatial division. In 
this sense, the entanglement entropy is considered to be a strong 
candidate for the statistical origin of the black hole entropy.

Now let us review the concept of the entanglement entropy. We consider 
a Hilbert space $\cal{F}$ constructed from two Hilbert spaces 
${\cal{F}}_{1}$ and ${\cal{F}}_{2}$ as 
%
\begin{equation}
 {\cal{F}} = {\cal{F}}_{1}\otimes{\cal{F}}_{2}.\label{eqn:F=F1*F2}
\end{equation}
From a pure density matrix $\rho=|\phi\rangle\langle\phi |$ on
$\cal{F}$ we can define reduced 
density matrices $\rho_{1}$ and $\rho_{2}$ by 
$\rho_{1,2}={\bf Tr}_{2,1}\rho$, 
where ${\bf Tr}_{1,2}$ represents a partial trace over 
${\cal{F}}_{1,2}$, respectively, and $|\phi\rangle$ is an element
of $\cal{F}$ with unit norm. The entanglement entropy is defined by 
$S_{ent1}=S[\rho_{1}]$ or $S_{ent2}=S[\rho_{2}]$, where 
$S[\cdot]$ is the von Neumann entropy. We can denote these two
entropies by the same symbol $S_{ent}$ since $S_{ent1}=S_{ent2}$ (see
Appendix of Ref.~\cite{MSK1} for a proof).

In the context of black hole thermodynamics the entanglement entropy
of matter fields on a black hole background is regarded as a strong
candidate for black hole entropy. In this case the Hilbert space
$\cal{F}$ is a space of all quantum states of matter fields and 
the direct product structure (\ref{eqn:F=F1*F2}) of $\cal{F}$ is
obtained from a division 
of a spacelike hypersurface into two regions: one inside and one
outside of a boundary surface. Due to the above mentioned symmetric
property $S_{ent}=S[\rho_{1}]=S[\rho_{2}]$, it is expected that
$S_{ent}$ for a pure state is proportional to the area $A$ of the
boundary. Thus, 
%
\begin{equation}
	S_{ent}=c\frac{A}{a^2}
\end{equation}
is expected, where $c$ is a numerical constant of 
order $1$ and $a$ is a cutoff length of the theory introduced in 
order to make the expression finite and dimensionless as entropy
should be. In many references~\cite{BKLS,Sred,Sent}, this behavior of 
$S_{ent}$ was confirmed. So the entanglement entropy for a pure state
has a property similar to black hole entropy, provided that the
boundary is set to be close to a black hole horizon and the cutoff
length is of Planck order. Moreover, in Ref.~\cite{MSK2}, it was shown
that concepts of entanglement energy and entanglement temperature can
be introduced for matter fields in a black hole background and that
their behavior is the same as for the energy and temperature of the
black hole. For these reasons the entanglement entropy has a potential
to be the origin of the black hole entropy.

In Sec.~\ref{sec:entropies}, based on a relation between the 
entanglement entropy and so-called conditional entropy, we propose an
interpretation of the entanglement entropy. In 
Sec.~\ref{sec:variational} variational principles in entanglement 
thermodynamics are used to determine quantum states. In particular, a 
state having maximal entanglement entropy is determined and is used 
in Sec.~\ref {sec:QT} to transmit information about an unknown
quantum state. Sec.~\ref{sec:summary} is devoted to a summary of this
paper and to discuss implications for the information loss problem and
Hawking radiation.


\section{Conditional entropy and entanglement entropy}
	\label{sec:entropies}

Entropy plays important roles not only in statistical mechanics but
also in information theory. In the latter, 
entropy of a random experiment, each of whose outcomes has an 
attached probability, represents uncertainty about the outcome before 
performing the experiment~\cite{Jumarie}. Besides the well-known 
Shannon entropy, there exist various definitions of entropies in 
information theory. For example, the so-called conditional 
entropy of an experiment $A$ on another experiment $B$ is defined by 
$H(A|B) = -\sum_{a,b}p(a,b)\ln p(a|b)$, where $a$ and $b$ represent 
outcomes of $A$ and $B$, respectively, 
$p(a,b)$ is a joint probability of $a$ and $b$, and $p(a|b)=p(a,b)/p(b)$ 
is a conditional probability of $a$ on $b$. Here $p(b)$ is a probability 
of $b$. The conditional entropy corresponds to an uncertainty about the
outcome of $A$ after the experiment $B$ is done. In other words it can 
be regarded as the amount of information about $A$ which cannot be 
known from the experiment $B$. The quantum analogue of the conditional 
entropy was considered in references~\cite{Cerf&Adami1,Cerf&Adami2}
and is called the von Neumann conditional entropy. Consider a Hilbert
space $\cal{F}$ of the form (\ref{eqn:F=F1*F2}) and let $\rho$ be a
density matrix on $\cal{F}$. The von Neumann conditional entropy of
$\rho$ about the subsystem $1$ on the subsystem $2$ is defined by 
%
\begin{equation}
	S_{1|2} = {\bf Tr}\left[\rho\sigma_{1|2}\right],
\end{equation}
where $\sigma_{1|2}={\bf 1}_{1}\otimes\ln\rho_{2}-\ln\rho$. The von 
Neumann conditional entropy $S_{2|1}$ of $\rho$ about the subsystem $2$ 
on the subsystem $1$ is defined in a similar way. It is expected that 
$S_{1|2}$ (or $S_{2|1}$) represents the amount of the information about 
the subsystem $1$ (or $2$) which cannot be known from $2$ (or 
$1$, respectively).

The von Neumann conditional entropy can be negative. In fact,
it is easy to see that 
%
\begin{equation}
	S_{1|2} = S_{2|1} = - S_{ent}, 
\end{equation}
if $\rho$ is a pure state. Hence, if $\rho$ is a pure state then
the conditional entropy is zero or negative. Our question now is `what
is the meaning of the negative 
conditional entropy of a pure state?' It might be expected that
$|S_{1|2}|$ ($=S_{ent}$) is the amount of the information about $1$
(or $2$) which can be known from $2$ (or $1$, respectively). 
However, this statement is not precise. A precise statement is that it 
is an amount of information, which can be transmitted through $1$ and
$2$ from a system interacting with $1$ to another system interacting
with $2$. The transmission medium is quantum entanglement
between $1$ and $2$.

The purpose of the remaining part of this paper is to give
suggestive arguments for this statement.


\section{Variational principles in entanglement thermodynamics}
	\label{sec:variational}
	
In statistical mechanics, the von Neumann entropy is used to determine 
an equilibrium state: an equilibrium state of an 
isolated system is determined by maximizing the entropy. Thus, we
expect that the entanglement entropy may be used to determine a 
quantum state.

As an illustration we consider a simple system of two 
particles, each with spin $1/2$: we consider a Hilbert space 
$\cal{F}$ of the form (\ref{eqn:F=F1*F2}) and denote an orthonormal 
basis of ${\cal{F}}_{i}$ by 
$\left\{ |\uparrow\rangle_{i},|\downarrow\rangle_{i}\right\}$ 
($i=1,2$). Let $|\phi\rangle$ be an element of $\cal{F}$ with unit 
norm and expand it as 
%
\begin{equation}
	|\phi\rangle = a|\uparrow\rangle_{1}\otimes|\uparrow\rangle_{2}
		+ b|\uparrow\rangle_{1}\otimes|\downarrow\rangle_{2}
		+ c|\downarrow\rangle_{1}\otimes|\uparrow\rangle_{2}
		+ d|\downarrow\rangle_{1}\otimes|\downarrow\rangle_{2},
		\label{eqn:stateinF}
\end{equation}
where $|a|^2+|b|^2+|c|^2+|d|^2=1$ is understood. The corresponding 
reduced density matrix is given by 
%
\begin{eqnarray}
 \rho_{2} & = &
 	(|a|^2+|c|^2)|\uparrow\rangle_{2}{}_{2}\langle\uparrow|
 	+ (ab^*+cd^*)|\uparrow\rangle_{2}{}_{2}\langle\downarrow|
 	\nonumber\\
 & & + (a^*b+c^*d)|\downarrow\rangle_{2}{}_{2}\langle\uparrow|
 	+ (|b|^2+|d|^2)|\downarrow\rangle_{2}{}_{2}\langle\downarrow|
\end{eqnarray}
and the entanglement entropy can be easily calculated from it. The resulting 
expression for the entanglement entropy is 
%
\begin{equation}
	S_{ent} = -\frac{1+x}{2}\ln\left(\frac{1+x}{2}\right)
		-\frac{1-x}{2}\ln\left(\frac{1-x}{2}\right),
\end{equation}
where $x=\sqrt{1-4|ad-bc|^2}$. By requiring $dS_{ent}/dx=0$ we 
obtain the condition $|ad-bc|=1/2$. Thus a state maximizing the 
entanglement entropy is
%
\begin{equation}
	|\phi\rangle = \frac{1}{\sqrt{2}}\left(
		|\uparrow\rangle_{1}\otimes|\downarrow\rangle_{2}
		-|\downarrow\rangle_{1}\otimes|\uparrow\rangle_{2}\right)
\end{equation}
up to a unitary transformation in ${\cal{F}}_{1}$ and the 
corresponding maximal value of the entanglement entropy is $\ln 2$. 
This state is well known as the EPR state.

It is notable that the 
corresponding reduced density matrix $\rho_{2}$
represents the microcanonical ensemble. This fact is related to the
fact that the maximum of entropy gives the microcanonical ensemble in
statistical mechanics. Thus, in general, if ${\cal{F}}_{1}$ and
${\cal{F}}_{2}$ have the same finite dimension $N$ then a state
maximizing the entanglement entropy is written as
%
\begin{equation}
	|\phi\rangle = \frac{1}{\sqrt{N}}\sum_{n=1}^N\left(
		|n\rangle_{1}\otimes|n\rangle_{2}\right)
		\label{eqn:EPRstate}
\end{equation}
up to a unitary transformation in ${\cal{F}}_{1}$, 
where $|n\rangle_{1}$ and $|n\rangle_{2}$ ($n=1,2,\cdots,N$) are 
orthonormal basis of ${\cal{F}}_{1}$ and ${\cal{F}}_{2}$, respectively. 
(See Appendix \ref{app:A} for a systematic derivation. )
In the next section we use the state (\ref{eqn:EPRstate}) to transmit
information about an unknown quantum state.

In statistical mechanics, free energy $F=E-TS$ can also be used to 
determine a statistical state: its minimum corresponds to an
equilibrium state of a subsystem in contact with a heat bath of
temperature $T$, provided that $T$ is fixed. This variational
principle in statistical mechanics is based on the following three
assumptions. 
\begin{enumerate}
\item The total system (the subsystem $+$ the heat bath) obeys the
principle of maximum of entropy.
\item Total energy (energy of the subsystem $+$ energy of the heat
bath) is conserved.
\item The 1st law of thermodynamics holds for the heat bath. 
\end{enumerate}
Is there a corresponding variational principle in 
the quantum system in the Hilbert space $\cal{F}$ of the form 
(\ref{eqn:F=F1*F2})? The answer is yes. In 
Ref.~\cite{MSK1} and Ref.~\cite{MSK2} a concept of entanglement energy 
was introduced and a thermodynamical structure, which we call 
entanglement thermodynamics, was constructed by using the entanglement 
entropy and the entanglement energy. Thus we expect that entanglement 
free energy $F_{ent}$ defined as follows plays an important role in 
entanglement thermodynamics.
%
\begin{equation}
	F_{ent} = E_{ent} - T_{ent}S_{ent},\label{eqn:Fent}
\end{equation}
where $E_{ent}$ is the entanglement energy and $T_{ent}$ is a 
constant. Among several options, we adopt the following definition of
the entanglement energy.
%
\begin{equation}
	E_{ent} = {\bf Tr}\left[\rho :H_{2}:\right],\label{eqn:Eent}
\end{equation}
where $H_{2}$ is Hamiltonian of the subsystem $2$, and $:-:$ represents 
the normal ordering~\cite{MSK2}.

As shown in the following arguments, by minimizing 
the entanglement free energy, we can obtain a state in $\cal{F}$ 
characterized by the constant $T_{ent}$. Before doing it, here we 
consider a physical meaning of the principle of minimum of the
entanglement free energy. Let us introduce another Hilbert space
${\cal{F}}_{bath}$, which plays the role of the heat bath in the above
statistical-mechanical consideration, and decompose it to the direct
product
${\cal{F}}_{bath}={\cal{F}}_{bath1}\otimes{\cal{F}}_{bath2}$. In this 
situation it is expected that the principle of the minimum entanglement 
free energy corresponds to the following situation.
\begin{enumerate}
\item The total system 
${\cal{F}}_{tot}\equiv{\cal{F}}\otimes{\cal{F}}_{bath}$ obeys the 
principle of  maximum of the entanglement entropy with respect to the 
decomposition 
${\cal{F}}_{tot}={\cal{F}}_{tot1}\otimes{\cal{F}}_{tot2}$, where
${\cal{F}}_{tot1}\equiv{\cal{F}}_{1}\otimes{\cal{F}}_{bath1}$ and 
${\cal{F}}_{tot2}\equiv{\cal{F}}_{2}\otimes{\cal{F}}_{bath2}$.
\item Total entanglement energy (entanglement energy for ${\cal{F}}$ 
$+$ entanglement energy for ${\cal{F}}_{bath}$) is conserved.
\item The 1st law of entanglement thermodynamics~\cite{MSK1}
%
\begin{equation}
 dE_{ent}=T_{ent}dS_{ent}
\end{equation}
holds for ${\cal{F}}_{bath}$. In this situation we call the constant
$T_{ent}$ the entanglement temperature. 
\end{enumerate}
It must be mentioned here that the variational principle of minimum of
the entanglement free energy is not as fundamental as the principle of
maximum of the entanglement entropy but is an approximation to the 
latter principle for a large system. However, like the principle of
minimum free energy in statistical mechanics, the former
principle should be a very useful tool to determine a quantum state.

We now calculate $F_{ent}$ for the system of two spin-$1/2$ particles 
and minimize it. For simplicity we adopt the following Hamiltonian for 
the subsystem $2$:
%
\begin{eqnarray}
	{}_{2}\langle\uparrow|:H_{2}:|\uparrow\rangle_{2} & = & 
		\epsilon,\nonumber\\
	{}_{2}\langle\uparrow|:H_{2}:|\downarrow\rangle_{2} & = & 
		0,\nonumber\\
	{}_{2}\langle\downarrow|:H_{2}:|\downarrow\rangle_{2} & = & 
		0,
\end{eqnarray}
where $\epsilon$ is a positive constant. The entanglement 
free energy $F_{ent}$ for the state (\ref{eqn:stateinF}) is given by
%
\begin{equation}
	F_{ent} = \epsilon(|a|^2 + |c|^2)
		+ T_{ent}\left[\frac{1+x}{2}\ln\left(\frac{1+x}{2}\right)
		+\frac{1-x}{2}\ln\left(\frac{1-x}{2}\right)\right].
\end{equation}
By minimizing it we obtain the following expression for the state 
$|\phi\rangle$ up to a unitary transformation in ${\cal{F}}_{1}$.
%
\begin{equation}
	|\phi\rangle = \frac{1}{\sqrt{Z}}\left[
		e^{-\epsilon/2T_{ent}}
		|\uparrow\rangle_{1}\otimes|\uparrow\rangle_{2}
		+ |\downarrow\rangle_{1}\otimes|\downarrow\rangle_{2}
		\right],
\end{equation}
where $Z=e^{-\epsilon/T_{ent}}+1$.

The corresponding reduced density matrix $\rho_{2}$ on ${\cal{F}}_{2}$
represents a canonical ensemble with temperature $T_{ent}$. This fact
is related to the fact that the principle of minimum of free energy
results in a canonical ensemble in statistical mechanics. Thus, in
general, if ${\cal{F}}_{1}$ and ${\cal{F}}_{2}$ have the same finite
dimension $N$ then a state minimizing the entanglement free
energy is written as 
%
\begin{equation}
	|\phi\rangle = \frac{1}{\sqrt{Z}}\sum_{n=1}^N\left[
		e^{-E_{n}/2T_{ent}}
		|n\rangle_{1}\otimes|n\rangle_{2}\right]
		\label{eqn:TFDstate}
\end{equation}
up to a unitary transformation in ${\cal{F}}_{1}$, where
$Z=\sum_{n=1}^Ne^{-E_{n}/T_{ent}}$, and $E_{n}$ and $|n\rangle_{2}$
($n=1,2,\cdots,N$) are eigenvalues and orthonormalized eigenstates of
the normal-ordered Hamiltonian of the subsystem $2$. (See Appendix
\ref{app:A} for a systematic derivation.)

The state (\ref{eqn:TFDstate}) can be
obtained also from another version of the principle of maximum of the 
entanglement entropy: if we maximize $S_{ent}$ with $E_{ent}$ fixed
then the state (\ref{eqn:TFDstate}) is obtained. In this case, the
constant $T_{ent}$ is determined so that the entanglement energy
coincides with the fixed value.

Note that in Eq. (\ref{eqn:TFDstate}) the infinite-dimensional limit
$N\to\infty$ can be taken, provided that $T_{ent}$ is bounded. 
In this limit, the state
(\ref{eqn:TFDstate}) has the same form as those appearing in 
the thermo field dynamics of black holes~\cite{Israel} and the quantum 
field theory on a collapsing star background~\cite{Parker}. In
fact, if we can set the value of the entanglement temperature of
${\cal{F}}_{bath}$ to be the black hole temperature then the state
(\ref{eqn:TFDstate}) in the limit completely coincides with those in 
Ref.~\cite{Israel,Parker}. In Ref.~\cite{MSK2} it was shown
numerically that the entanglement temperature for a real 
massless scalar field in a Schwarzschild spacetime is finite and equal
to the black hole temperature of the background geometry up to a
numerical constant of order $1$. The finiteness of the entanglement
temperature in the Schwarzschild spacetime is a result of 
cancellation of divergences in entanglement entropy and entanglement
energy~\cite{MSK2}. Thus, the finiteness is preserved even in the
limit of zero cutoff length ($a\to 0$).


\section{Quantum teleportation}
	\label{sec:QT}

In Ref.~\cite{Bennett-etal} Bennet et al. proposed a method of 
teleportation of an unknown quantum state from one place to another. It 
is called quantum teleportation. In their method the information
about the quantum state is separated into a `quantum channel' and a
`classical channel', and each channel is sent separately from a sender
``Alice'' to a receiver ``Bob''. What is important is that the quantum
channel is sent in a superluminal way by using a quantum correlation
or entanglement, while the classical channel is transmitted at most in
the speed of light. Here we mention that causality is not violated in
an informational sense since Bob cannot obtain any useful information
about the unknown state before the arrival of the classical
channel. Hence Alice has to deliver the classical channel to Bob
without fail. On the contrary she
does not need to worry about whether the information in the quantum
channel arrives at Bob's hand since the arrival is guaranteed by the
quantum mechanics. It is notable that recently quantum
teleportation was confirmed by
experiments~\cite{Bouwrneester-etal,Boschi-etal}.

In this section we generalize the arguments in
Ref.~\cite{Bennett-etal} to more abundant situations and try to
reformulate it in terms of the entanglement entropy.

Let us consider a Hilbert space ${\cal{F}}$ of the form 
(\ref{eqn:F=F1*F2}) with ${\cal{F}}_{i}$ constructed from Hilbert 
spaces ${\cal{F}}_{i\pm}$ as
%
\begin{equation}
	{\cal{F}}_{i} = {\cal{F}}_{i+}\otimes{\cal{F}}_{i-}. 
	\label{Fi=Fi+*Fi-}
\end{equation}
For example, consider matter fields in a black hole spacetime formed by 
gravitational collapse. In this situation, let ${\cal{H}}_{1}$ be a space 
of all wave packets on the future event horizon and ${\cal{H}}_{2}$ be 
a space of all wave packets on the future null infinity, and decompose 
each ${\cal{H}}_{i}$ into a high frequency part ${\cal{H}}_{i+}$ and a 
low frequency part ${\cal{H}}_{i-}$. Typically, we suppose the 
decomposition at an energy scale of Planck order. If we define 
${\cal{F}}_{i\pm}$ as Fock spaces constructed from ${\cal{H}}_{i\pm}$, 
respectively, then the space $\cal{F}$ of all quantum states of the matter 
fields is given by (\ref{eqn:F=F1*F2}) with (\ref{Fi=Fi+*Fi-}). 
Although the following arguments do not depend on the construction 
of the Hilbert space $\cal{F}$, this example should be helpful for us
to understand the physical meaning of the results obtained.

For simplicity we consider the case 
that all ${\cal{F}}_{i\pm}$ have the same finite dimension $N$ 
although in the above example of field theory the dimensions of the
Hilbert spaces are infinite~\footnote{
In applying the results for finite dimensions to field theory,
we have to introduce a regularization scheme to make the system
finite. For example, we can discretize the system by introducing a
cutoff length. After that, we can consider a finite dimensional
subspace of the total Hilbert space of the discretized theory, for
example, by restricting total energy to be less than the mass of the
background geometry. After performing all calculations, we have to
confirm that the infinite-dimensional limit can be taken. See, for
example, the final paragraph of the previous section.
}. 
The main purpose of this
section is to see general properties of the entanglement
entropy by using a finite system. Anyway, the above example of field
theory may be helpful to understand the following arguments. 
It will be investigated elsewhere how to generalize the arguments of
quantum teleportation to infinite dimensions. 
In the finite dimensional case we assume the following three physical
principles. 
\begin{enumerate}
 \renewcommand{\labelenumi}{(\alph{enumi})}
\item A quantum state $|\phi\rangle$ in $\cal{F}$ is a direct product 
state given by $|\phi_{+}\rangle_{+}\otimes|\phi_{-}\rangle_{-}$, 
where $|\phi_{\pm}\rangle_{\pm}$ are elements of 
${\cal{F}}_{\pm}={\cal{F}}_{1\pm}\otimes{\cal{F}}_{2\pm}$, respectively.
\item $|\phi_{+}\rangle_{+}$ is determined by the principle of
maximum of the entanglement entropy with respect to the decomposition 
${\cal{F}}_{+}={\cal{F}}_{1+}\otimes{\cal{F}}_{2+}$.
\item A complete measurement of the von Neumann type on the joint 
system ${\cal{F}}_{1}$ is performed by a sender (Alice) in the
orthonormal basis $\{|\psi_{nm}\rangle_{1}\}$, each of 
which maximizes the entanglement entropy with respect to the 
decomposition ${\cal{F}}_{1}={\cal{F}}_{1+}\otimes{\cal{F}}_{1-}$. 
\end{enumerate}
In other words the assumption (c) is stated as follows: 
the state $|\phi\rangle$ is projected by one of the basis 
$|\psi_{nm}\rangle_{1}$.

In the following arguments, under these 
assumptions, we show a possibility of quantum teleportation of the 
state $|\phi_{-}\rangle_{-}$ in ${\cal{F}}_{-}$ to ${\cal{F}}_{2}$: we 
make a clone of $|\phi_{-}\rangle_{-}$ by using the quantum
entanglement which the state $|\phi_{+}\rangle_{+}$ has. 
Therefore a receiver (Bob), who cannot contact with ${\cal{F}}_{1}$,
may be able to get all information about the state
$|\phi_{-}\rangle_{-}$ in ${\cal{F}}_{-}$, provided that he can manage
to get the classical channel.

Now let us show that explicitly. By the assumption (b) and the
arguments in Sec.~\ref{sec:variational} (see
Eq.~(\ref{eqn:EPRstate})), the state $|\phi_{+}\rangle_{+}$ can be
written as 
%
\begin{equation}
	|\phi_{+}\rangle_{+} = \frac{1}{\sqrt{N}}\sum_{n=1}^N
		|n\rangle_{1+}\otimes|n\rangle_{2+},\label{eqn:phi+}
\end{equation}
where $\{|n\rangle_{i+}\}$ ($n=1,2,\cdots,N$) are an orthonormal basis 
of ${\cal{F}}_{i+}$. Next, expand $|\phi_{-}\rangle_{-}$ as
%
\begin{equation}
	|\phi_{-}\rangle_{-} = \sum_{nm}C_{nm}
		|n\rangle_{1-}\otimes|m\rangle_{2-},\label{eqn:phi-}
\end{equation}
where $\{|n\rangle_{i-}\}$ ($n=1,2,\cdots,N$) are an orthonormal basis 
of ${\cal{F}}_{i-}$, and $\sum_{nm}|C_{nm}|^2=1$ is understood. 
To impose the assumption (c), we adopt the following 
basis $\{|\psi_{nm}\rangle_{1}\}$ ($n,m=1,2,\cdots,N$), each of which 
maximizes the entanglement entropy.
%
\begin{equation}
 |\psi_{nm}\rangle_{1} = \frac{1}{\sqrt{N}}\sum_{j=1}^N 
 e^{2\pi ijn/N}|(j+m)modN\rangle_{1+}\otimes|j\rangle_{1-}.
 	\label{eqn:BELLstates}
 \end{equation}
In Appendix \ref{app:B} it is proved that (\ref{eqn:BELLstates}) is
unique up to a unitary transformation in ${\cal{F}}_{1+}$. 
Hence, $|\phi\rangle=|\phi_{+}\rangle_{+}\otimes|\phi_{-}\rangle_{-}$ is 
written as 
%
\begin{equation}
 |\phi\rangle = \frac{1}{N}\sum_{nm}|\psi_{nm}\rangle_{1}\otimes
		U^{(2+)}_{nm}|\tilde{\phi}_{2}\rangle_{2},
		\label{eqn:phi-decomp}
\end{equation}
where $|\tilde{\phi}_{2}\rangle_{2}$ is a state in ${\cal{F}}_{2}$ 
given by 
%
\begin{equation}
 |\tilde{\phi}_{2}\rangle_{2} = 
	\sum_{n'm'}C_{n'm'}|n'\rangle_{2+}\otimes|m'\rangle_{2-},
	\label{eqn:tilde-phi2}
\end{equation}
and $U^{(2+)}_{nm}$ $(n,m=1,2,\cdots,N$) are unitary transformations 
in ${\cal{F}}_{2+}$ defined by 
%
\begin{equation}
 U^{(2+)}_{nm} = \sum_{k=1}^N 
	e^{-2\pi ikn/N}|(k+m)modN\rangle_{2+}{}_{2+}\langle k|.
	\label{eqn:Unitary-tr}
 \end{equation}
(See Appendix \ref{app:B} for an explicit derivation of
(\ref{eqn:phi-decomp}).)

Thus, after the measurements in the basis $\{|\psi_{nm}\rangle_{1}\}$ 
by the sender (Alice), the original state $|\phi\rangle$ jumps to one
of the states $|\tilde{\phi}_{nm}\rangle$ defined by 
%
\begin{equation}
 |\tilde{\phi}_{nm}\rangle = |\psi_{nm}\rangle_{1}\otimes
		U^{(2+)}_{nm}|\tilde{\phi}_{2}\rangle_{2}. 
		\label{eqn:tild-phi}
\end{equation}
This state can be seen by the receiver (Bob), who cannot contact 
with ${\cal{F}}_{1}$, as the state 
$U^{(2+)}_{nm}|\tilde{\phi}_{2}\rangle_{2}$ in ${\cal{F}}_{2}$. Here 
note that the unitary transformation $U^{(2+)}_{nm}$ in 
${\cal{F}}_{2+}$ is completely determined by a pair of integers $n$ and 
$m$ (outcome of the experiment by Alice). Thus, if the two integers 
are sent to the receiver (Bob) in the classical channel, then by
operating the inverse transformation of the corresponding unitary
transformation in ${\cal{F}}_{2+}$ the receiver (Bob) can obtain the
`clone' state $|\tilde{\phi}_{2}\rangle_{2}$ ($\in {\cal F}_{2}$) of
$|\phi_{-}\rangle$ ($\in {\cal F}_{-}$). It is evident that
$|\tilde{\phi}_{2}\rangle_{2}$ has all information about the original 
state $|\phi_{-}\rangle$.

It is remarkable that information to be 
sent to the receiver (Bob) in the classical channel is only two 
integers $n$ and $m$, while information included in the unknown state
$|\phi_-\rangle_-$ is a set of complex constants $\{C_{nm}\}$
($n,m=1,2,\cdots,N$) with a constraint $\sum_{nm}|C_{nm}|^2=1$. Thus a 
large amount of information is sent in the quantum channel. 
Here we mention that tracing out ${\cal{F}}_{2+}$ from the state 
$U^{(2+)}_{nm}|\tilde{\phi}_{2}\rangle_{2}$ or
$|\tilde{\phi}_{2}\rangle_{2}$ results in the following density matrix
$\rho_{2-}$ on ${\cal{F}}_{2-}$: 
%
\begin{equation}
 \rho_{2-} = \sum_{nm}\left(\sum_{j}C_{jn}C_{jm}^*\right)
 	|n\rangle_{2-}{}_{2-}\langle m|, \label{eqn:rho2-}
\end{equation}
which is equivalent to the density matrix obtained by tracing out 
${\cal{F}}_{1-}$ from the original unknown state
$|\phi_{-}\rangle_{-}$. Hence, if the receiver (Bob) cannot contact
with ${\cal F}_{2+}$, he does not obtain any information from the
sender (Alice).

Finally it must be mentioned that the success of quantum 
teleportation is due to the state $|\phi_{+}\rangle_{+}$ which has 
maximal entanglement entropy. If we took $|\phi_{+}\rangle_{+}$ 
with less entanglement entropy then the teleportation would be less 
successful. Therefore, the entanglement entropy can be regarded as an 
index of efficiency of quantum teleportation. This consideration 
supports the interpretation of the entanglement entropy proposed in 
Sec.~\ref{sec:entropies}.


\section{Summary and discussion}
	\label{sec:summary}

In this paper a new interpretation of entanglement entropy has been
proposed based on its relation to the so-called conditional entropy and 
a well-known meaning of the latter. It is conjectured that 
entanglement entropy of a pure state with respect to a division of a 
Hilbert space into two subspaces $1$ and $2$ is an amount of
information, which can be transmitted through $1$ and $2$ from a
system interacting with $1$ to another system interacting with
$2$. The medium of the transmission is quantum entanglement between
$1$ and $2$.

To support the interpretation we have given the following two suggestive
arguments: variational principles in entanglement thermodynamics and
quantum teleportation. The most important variational principle we
considered is the principle of maximum of entanglement entropy. This
principle determines a state uniquely up to a unitary transformation
in one of the two Hilbert subspaces (not in the whole Hilbert
space). From the proposed conjecture it is expected that information
can be transmitted most effectively through the two subspaces by using 
the maximal entanglement of the state. 
In fact, reformulating the quantum teleportation in terms of the
entanglement entropy, we have shown that the state having maximal
entanglement entropy plays an important role in quantum
teleportation. This consideration gives strong support to our
interpretation.

It is confirmed in many references~\cite{BKLS,Sred,Sent} that the 
entanglement entropy has the same value as the black hole entropy up to 
a numerical constant of order $1$, provided that a cutoff length of
Planck order is introduced in the theory. Hence we have a large 
amount of entanglement entropy to transmit information from
inside to outside of a black hole by using quantum entanglement.

Hence, in our interpretation, it seems that the
entanglement entropy is a quantity which cancels the black hole
entropy to restore information loss, provided that the black hole
entropy represents the amount of the information loss. 
For example, suppose that a black hole is formed from 
an initial state with zero entropy ($S=0$). In this case, 
non-zero black hole entropy is generated ($S_{BH}>0$) from the zero
entropy state. At the same time, entanglement entropy and negative
conditional entropy are also generated and their absolute values are as
large as the black hole entropy ($S_{ent}=|S_{cond}|\simeq S_{BH}$). 
After that, the black hole evolves by emitting Hawking 
radiation, changing the value of $S_{BH}$ and $S_{ent}$
($=|S_{cond}|$) with $S_{BH}\simeq S_{ent}$ kept. Finally, when the
black hole evaporates, the entanglement entropy cancels the
black hole entropy to settle the final entropy to be zero
($S=0$). To summarize, the black hole entropy is an amount of
temporarily missing information and the entanglement entropy is a
quantity which cancels the black hole entropy. Both entropies appear
and disappear together from the sea of zero entropy state.

As a by-product we have shown that the variational principle of 
minimum of entanglement free energy is useful to determine a quantum 
state. The resulting quantum state has exactly the same form as those 
appearing in the thermo field dynamics of black holes~\cite{Israel}
and the quantum field theory on a collapsing black hole
background~\cite{Parker}, provided that the entanglement temperature
$T_{ent}$ is set to be the black 
hole temperature. It is remarkable that $T_{ent}$ for a real massless scalar 
field in a Schwarzschild spacetime is equal to the black hole temperature 
of the background geometry up to a numerical constant of order 
$1$~\cite{MSK2}. Thus we can say that the variational principle of 
minimum of entanglement free energy gives a new derivation of the 
Hawking radiation. 
Finally, we mention that with this variational principle the
entanglement thermodynamics is equivalent to 'tHooft's brick wall
model~\cite{Brick-Wall}.

It will be valuable to analyze how to generalize arguments in this
paper to the situation that divergences in entropy and energy are
absorbed by renormalization~\cite{Sreg,induced-G}. 
If the generalization is achieved, the physical meaning of the 
entanglement entropy in black hole physics will become clearer. 
It is noteworthy that in the brick wall model the divergence in 
thermal energy is exactly canceled by divergence in negative 
energy~\cite{Mukohyama&Israel}.

Now the final comment is in order. It is worthwhile to clarify in what 
physical situations the variational principles can be applicable. (In 
thermodynamics the second law supports the principle of maximum 
entropy.) In other words, in what situations does the entanglement 
entropy increase? 
In what situations does the entanglement free energy decrease? 
To answer these questions, theorem 2 of Ref.~\cite{Mukohyama-GSL} or 
its generalization may be useful.

\vskip 1cm

\centerline{\bf Acknowledgments}
The author thanks Professor W. Israel for helpful discussions and so
many stimulating suggestions. Without his help, this work could not be 
done. The author is grateful to S. A. Hayward for a careful reading of 
the manuscript and to Professor H. Kodama for continuous
encouragement. 
The author was supported by JSPS Research Fellowships for Young 
Scientists, and this work was supported partially by the Grand-in-Aid
for Scientific Research Fund (No. 9809228).

\appendix

\section{}
	\label{app:A}

In this appendix we give derivations of formulas (\ref{eqn:EPRstate})
and (\ref{eqn:TFDstate}).

We consider a Hilbert space $\cal{F}$ of the form
%
\begin{equation}
 {\cal{F}} = {\cal{F}}_{1}\otimes{\cal{F}}_{2},
\end{equation}
where ${\cal{F}}_{1}$ and ${\cal{F}}_{2}$ are Hilbert spaces with the
same finite dimension $N$. An arbitrary unit element $|\phi\rangle$ of 
$\cal{F}$ is decomposed as 
%
\begin{equation}
 |\phi\rangle = \sum_{n=1}^N \sum_{m=1}^N 
	C_{nm}|n\rangle_1 \otimes |m\rangle_2,
\end{equation}
where $|n\rangle_{1}$ and $|n\rangle_{2}$ ($n=1,2,\cdots,N$) are 
orthonormal bases of ${\cal{F}}_{1}$ and ${\cal{F}}_{2}$,
respectively, and $\sum_{n,m}|C_{nm}|^2=1$ is
understood. Here, without loss of generality, we can choose the
orthonormal basis of ${\cal{F}}_{2}$ to be eigenstates of
the normal-ordered Hamiltonian $:H_2:$ of the sub-system as
%
\begin{equation}
 :H_2: |n\rangle_2 = E_n |n\rangle_2. 
\end{equation}
Since $C^{\dagger}C$ is a non-negative hermitian matrix, we can define a
set of non-negative real numbers $\{ p_n\}$, each of which is the 
eigenvalue of the matrix $C^{\dagger}C$. Hence,
%
\begin{equation}
 C^{\dagger}C = V^{\dagger}PV,
\end{equation}
where $V$ is a unitary matrix and $P$ is a diagonal matrix with
diagonal elements $\{ p_n\}$. With these definitions, the entanglement
entropy $S_{ent}$ and the entanglement free energy $F_{ent}$ are
calculated as 
%
\begin{eqnarray}
 S_{ent} & = & -\sum_{n=1}^N p_n\ln p_n,\label{eqn:Sent-pn}\\
 F_{ent} & = & \sum_{n=1}^N\sum_{m=1}^N E_n p_m |V_{mn}|^2
		+ T_{ent}\sum_{n=1}^N p_n\ln p_n. 
		\label{eqn:Fent-pn}
\end{eqnarray}
The constraints $\sum_{n,m}|C_{nm}|^2=1$ and $V^{\dagger}V={\bf 1}$
are equivalent to 
%
\begin{eqnarray}
 \sum_{n=1}^N p_n & = & 1,\nonumber\\
 \sum_{l=1}^N V^{*}_{ln}V_{lm} & = & \delta_{nm}.
		\label{eqn:constraint}
\end{eqnarray}
Thus, the variational principles are restated as follows: to maximize
(\ref{eqn:Sent-pn}) under the constraints (\ref{eqn:constraint}); to
minimize (\ref{eqn:Fent-pn}) under the constraints
(\ref{eqn:constraint}).

Now, we shall show that expressions (\ref{eqn:Sent-pn}) and
(\ref{eqn:Fent-pn}) are same as those for entropy and free energy
in statistical mechanics in the subspace ${\cal{F}}_2$.  Let us consider
a density operator $\bar{\rho}$ on ${\cal{F}}_2$: 
%
\begin{equation}
 \bar{\rho} = \sum_{n=1}^N \sum_{m=1}^N \tilde{P}_{nm}
		|n\rangle_{2}{}_{2}\langle m|,
\end{equation}
where $\tilde{P}_{nm}$ is an $N\times N$ non-negative hermitian matrix
with unit trace.  By diagonalizing the matrix $\tilde{P}$ as
%
\begin{equation}
 \tilde{P} = \bar{V}^{\dagger}\bar{P}\bar{V},
\end{equation}
we obtain the following expressions for entropy and free energy.
%
\begin{eqnarray}
 S & = & -\sum_{n=1}^N \bar{p}_n\ln \bar{p}_n,\nonumber\\
 F & = & \sum_{n=1}^N\sum_{m=1}^N E_n \bar{p}_m |V_{mn}|^2
		+ T\sum_{n=1}^N \bar{p}_n\ln \bar{p}_n,
\end{eqnarray}
where $\{\bar{p}_n\}$ are diagonal elements of the matrix $\bar{P}$ 
and $T$ is temperature. The constraints $\bf{Tr}\bar{\rho}=1$  and
$\bar{V}^{\dagger}\bar{V}={\bf 1}$ are restated as 
%
\begin{eqnarray}
 \sum_{n=1}^N \bar{p}_n & = & 1,\nonumber\\
 \sum_{l=1}^N \bar{V}^{*}_{ln}\bar{V}_{lm} & = & \delta_{nm}.
\end{eqnarray}

At this point, it is evident that the variational principles of
maximum of entropy are the same in entanglement thermodynamics and
statistical mechanics and that the principles of minimum of free
energy are also the same in the two schemes. Hence, the principle of
maximum of the entanglement entropy gives 
%
\begin{equation}
 \left(C^{\dagger}C\right)_{nm} = \frac{1}{N}\delta_{nm}, 
		\label{eqn:minSent}
\end{equation}
as the principle of maximum of entropy gives the microcanonical
ensemble 
%
\begin{equation}
 \tilde{P}_{nm} = \frac{1}{N}\delta_{nm}
\end{equation}
in statistical mechanics. Similarly, the principle of minimum of
the entanglement free energy gives
%
\begin{equation}
 \left( C^{\dagger}C\right)_{nm} = 
		Z^{-1}e^{-E_n/T_{ent}}\delta_{nm},
			\label{eqn:maxFent}
\end{equation}
as the principle of minimum of the free energy gives the canonical
ensemble 
%
\begin{equation}
 \tilde{P}_{nm} = \bar{Z}^{-1}e^{-E_n/T}\delta_{nm}
\end{equation}
in statistical mechanics, where $Z=\sum_ne^{-E_n/T_{ent}}$ and
$\bar{Z}=\sum_ne^{-E_n/T}$ . It is easy 
to see that (\ref{eqn:minSent}) and (\ref{eqn:maxFent}) are equivalent 
to (\ref{eqn:EPRstate}) and (\ref{eqn:TFDstate}), respectively, up to
a unitary transformation in ${\cal{F}}_1$.

Finally we comment on the generalization of the analysis when the 
Hilbert space is divided into two subspaces with different 
dimensions ($dim{\cal F}_{1}>dim{\cal F}_{2}$). 
In this case, by defining $S_{ent}$ from $\rho_{2}$, we obtain 
similar results.

\section{}
	\label{app:B}

In this appendix we show that in the finite dimensional case the
orthonormal basis $\{|\psi_{nm}\rangle_{1}\}$ defined in the physical
principle (c) is given by (\ref{eqn:BELLstates}) uniquely up to a
unitary transformation in ${\cal{F}}_{1+}$. After that, we derive the
equation (\ref{eqn:phi-decomp}).

We consider the following decomposition of the Hilbert space
${\cal{F}}_{1}$. 
%
\begin{equation}
 {\cal{F}}_{1}={\cal{F}}_{1+}\otimes{\cal{F}}_{1-},
\end{equation}
where ${\cal{F}}_{1+}$ and ${\cal{F}}_{1-}$ are Hilbert spaces with
the same finite dimension $N$. From the arguments in Appendix
\ref{app:A}, each of the basis $\{|\psi_{nm}\rangle_{1}\}$ is obtained
by applying a unitary transformation in ${\cal{F}}_{1+}$ to the
following state in ${\cal{F}}_1$. 
%
\begin{equation}
 |\phi\rangle_1 = \frac{1}{\sqrt{N}}\sum_{j=1}^N\left(
		|j\rangle_{1+}\otimes|j\rangle_{1-}\right),
\end{equation}
where $|j\rangle_{1+}$ and $|j\rangle_{1-}$ ($j=1,2,\cdots,N$) are 
orthonormal bases of ${\cal{F}}_{1+}$ and ${\cal{F}}_{1-}$,
respectively. Evidently, any states given by (\ref{eqn:BELLstates})
are obtained by this procedure. Moreover, it is easily confirmed as
follows that a set of all states given by (\ref{eqn:BELLstates}) is a
complete orthonormal basis in the $N\times N$ dimensional Hilbert
space ${\cal{F}}_1$. 
%
\begin{equation}
 {}_{1}\langle\psi_{nm}|\psi_{n'm'}\rangle_{1} =
	\frac{1}{N}\sum_{j=1}^{N} e^{2\pi ij(n'-n)/N}\delta_{mm'}
	= \delta_{nn'}\delta_{mm'}.
\end{equation}

Let us suppose another complete orthonormal basis
$\{|\bar{\psi}_{nm}\rangle_{1}\}$ in ${\cal{F}}_1$, each of which
maximizes the entanglement entropy with respect to the decomposition
${\cal{F}}_{1}={\cal{F}}_{1+}\otimes{\cal{F}}_{1-}$.
Since both $\{|\psi_{nm}\rangle_{1}\}$ and
$\{|\bar{\psi}_{nm}\rangle_{1}\}$ are complete orthonormal basis in
${\cal{F}}_1$, they are related by a unitary transformation $U$ in
${\cal{F}}_1$. Moreover, $U$ is a unitary transformation in
${\cal{F}}_{1+}$, since any states maximizing the entanglement entropy 
are related by unitary transformations in ${\cal{F}}_{1+}$ as
shown in Appendix \ref{app:A}. Therefore, the orthonormal basis 
$\{|\psi_{nm}\rangle_{1}\}$ defined in the physical principle (c) is
unique up to a unitary transformation in ${\cal{F}}_{1-}$ and is
given by (\ref{eqn:BELLstates}).

Now let us show the equation (\ref{eqn:phi-decomp}). The right hand
side is transformed as follows.
%
\begin{eqnarray}
 \frac{1}{N}\sum_{nm}|\psi_{nm}\rangle_{1}\otimes
		U^{(2+)}_{nm}|\tilde{\phi}_{2}\rangle_{2}
 & = & \frac{1}{\sqrt{N}}\sum_{jk}
	\left(\frac{1}{N}\sum_{n}e^{2\pi i(j-k)n/N}\right)\times
	\sum_{mm'n'}{}_{2+}\langle k|n'\rangle_{2+} C_{n'm'}
	\nonumber\\
 & & 	\times |(j+m)modN\rangle_{1+}\otimes |j\rangle_{1-}
	\otimes|(k+m)modN\rangle_{2+}\otimes |m'\rangle_{2-}
	\nonumber\\
 & = &	\sum_{mm'n'}\left(\frac{1}{\sqrt{N}}|(n'+m)modN\rangle_{1+}
	\otimes |(n'+m)modN\rangle_{2+}\right) 
	\nonumber\\
 & &	\times 	C_{n'm'}|n'\rangle_{1-}\otimes|m'\rangle_{2-}
	\nonumber\\
 & = &	\left(\frac{1}{\sqrt{N}}\sum_{m''}|m''\rangle_{1+}
	\otimes |m''\rangle_{2+}\right) 
	\nonumber\\
 & &	\times \left(\sum_{n'm'}C_{n'm'}
	|n'\rangle_{1-}\otimes|m'\rangle_{2-}\right).
\end{eqnarray}
The final expression is $|\phi\rangle$ itself.


\end{document}